\documentclass[preprint]{aastex7}

\hypersetup{linkcolor=red,citecolor=blue,filecolor=cyan,urlcolor=violet}

\usepackage{graphicx}
\usepackage{amsmath}
\usepackage{graphicx}
\graphicspath{{bilder/}}
\usepackage{bbm}

\providecommand{\apjl}{\textit{Astrophysical Journal Letters}}

\begin{document}

\title{Heat-flux instabilities of regularized Kappa distributed strahl electrons resolved with ALPS}

\author[0009-0007-7679-5275]{Dustin L. Schröder}
\affiliation{Institut für Theoretische Physik, Lehrstuhl IV: Plasma-Astroteilchenphysik, Ruhr-Universität Bochum, D-44780 Bochum,
Germany}
\email[show]{dustin.schroeder@rub.de}  

\author[0000-0002-8508-5466]{Marian Lazar}
\affiliation{Institut für Theoretische Physik, Lehrstuhl IV: Plasma-Astroteilchenphysik, Ruhr-Universität Bochum, D-44780 Bochum,
Germany}
\affiliation{Centre for mathematical Plasma-Astrophysics, KU Leuven, 3001 Leuven, Belgium}
\email{marianlazar@yahoo.com} 

\author[0000-0003-3223-1498]{Rodrigo A. López}
\affiliation{Research Center in the intersection of Plasma Physics, Matter, and Complexity ($P^2 mc$),\\ Comisi\'on Chilena de Energ\'{\i}a Nuclear, Casilla 188-D, Santiago, Chile}
\affiliation{Departamento de Ciencias F\'{\i}sicas, Facultad de Ciencias Exactas, Universidad Andres Bello, Sazi\'e 2212, Santiago 8370136, Chile}
\email{rodrigo.lopez@cchen.cl} 

\author[0000-0002-9151-5127]{Horst Fichtner}
\affiliation{Institut für Theoretische Physik, Lehrstuhl IV: Plasma-Astroteilchenphysik, Ruhr-Universität Bochum, D-44780 Bochum,
Germany}
\email{hf@tp4.rub.de}



\begin{abstract}

The fluid behavior of the solar wind is affected by the heat flux carried by the suprathermal electron populations, especially the electron strahl (or beam) that propagates along the magnetic field. 
In turn, the electron strahl cannot be stable, and in the absence of collisions, its properties are regulated mainly by self-generated instabilities.
This paper approaches the description of these heat-flux instabilities in a novel manner using regularized Kappa distributions (RKDs) to characterize the electron strahl. 
RKDs conform to the velocity distributions with suprathermal tails observed in-situ, and at the same time allow for consistent macromodeling, based on their singularity-free moments.
In contrast, the complexity of RKD models makes the analytical kinetic formalism complicated and still inaccessible, and therefore, here heat-flux instabilities are resolved using the advanced solver ALPS. 
Two primary types of instabilities emerge depending on plasma conditions: the whistler and firehose heat-flux instabilities.
The solver is successfully tested for the first time for such instabilities by comparison with previous results for standard distributions, such as Maxwellian and Kappa.
Moreover, the new RKD results show that idealized Maxwellian models can overrate or underestimate the effects of these instabilities, and also show differences from those obtained for the standard Kappa, which, for instance, underestimate the firehose heat-flux growth rates.


\end{abstract}



\section{Introduction}\label{sec:intro}

The last decade has marked important progress in the development of numerical tools for the kinetic calculation of plasma waves and microinstabilities \citep{Astfalk-etal-2015, Astfalk-Jenko-2017, Verscharen_2018, Lopez_Shaaban_Lazar_2021,  rodrigo_a_lopez_DISK, ALPS:2023}. 
New numerical capabilities and techniques have been developed to interpret the wave dispersion and stability properties of plasma populations with nonequilibrium velocity distributions that depart from idealized Maxwellian models, such as Kappa power-law distributions reported by the observations in space plasmas. 
Thus, it should be mentioned the advanced 3D expressions of the dielectric tensors for plasmas with Kappa distributions that have been derived to account for various kinetic anisotropies, such as beams and temperature anisotropies \citep{Liu-etal-2014, Gaelzer-etal-2016a, Gaelzer-etal-2016b, Kim_2017}, and to facilitate the building of new specific numerical solvers that can resolve the full spectra of stable or unstable waves \citep{Astfalk-etal-2015, Lopez_Shaaban_Lazar_2021}.
Introduced as empirical models to reproduce the observed distributions with suprathermal tails \citep{Olbert1968, Vasyliunas1968}, the standard Kappa distributions (SKD) nevertheless suffer from the unphysical contributions of superluminal particles of the enhanced tails, and the limitations imposed on the power exponent $\kappa$ (for the consistency of higher-order moments which kinetically define macroparameters, such as temperature or pressure and heat flux) \citep{Scherer2018, Scherer_2019a, Lazar-Fichtner-2021}. 
These were overcome by adjusting the non-equilibrium models to what is already known as the regularized Kappa distribution (RKD), which suppresses the unphysical intake and allows for the consistent definition of all moments for any values of the parameter $\kappa>0$ \citep{Scherer2018, Scherer_2019a, Scherer_2019}.

However, the analysis of waves and instabilities requires the derivation of the dielectric response of plasmas with such distributions, which is not a straightforward task. So far, only facile expressions for electrostatic excitations in the absence of a magnetic field have been obtained \citep{Scherer2018, Gaelzer-Fichtner2024}.
Instead, here we exploit a new generation numerical solver called ALPS, the Arbitrary Linear Plasma Solver \citep{Verscharen_2018, ALPS:2023}, which does not require {\it a priori} derivation of these dielectric properties specific to RKDs.
We do this for the first time for the heat-flux instabilities of the strahl (or beam) population of electrons observed in the solar wind \citep{Wilson-etal-2019a, Wilson-etal-2019b, Bercic-etal-2019, Abraham-etal-2022}. 
The suprathermal electron strahl is the main carrier of heat flux in the solar wind, regulated most probably by self-generated instabilities \citep{Shaaban_2018MNRAS, Shaaban_2018PoP, Shaaban_2019MNRAS, Verscharen-etal-2019, Lopez-etal-2020, Micera-et-al-2020, Zenteno-Quinteros-et-al-2021}.
Evaluations of this heat flux for SKD or RKD strahls can be found in \cite{Lazar-etal-2020}.
The results presented here continue the series of recent works \citep{Husidic-etal-ASS-2021, SchröderPoP2025} aiming to support advanced numerical tools capable of directly solving the properties of plasmas with arbitrary distributions \citep{Astfalk-Jenko-2017, Verscharen_2018, ALPS:2023}.
Their application potential is huge in the context of the progress made by numerical resources and the demand for real-time analysis of data measured in situ in space plasmas.

The present paper is structured as follows. 
Section \ref{sec:theory} provides a concise overview of the theoretical framework for a dual-core–strahl electron modeling, followed by a description of the wave dispersion formalism used to study wave instabilities as implemented in ALPS. 
In Section \ref{sec:validation}, we validate the ALPS implementation through comparison with established results, with particular emphasis on the studies by \cite{Shaaban_2018MNRAS, Shaaban_2018PoP} which address parallel heat-flux unstable modes in electron plasmas composed of a Maxwellian core and an SKD-distributed strahl.
The heat-flux instabilities (HFIs) under consideration primarily include the whistler HFI (WHFI) and firehose HFI (FHFI), both driven by counterstreaming electron populations. 
The WHFI, characterized by right-hand (RH) polarization, is specific to sufficiently low beam velocities, whereas the left-hand (LH) polarized FHFI is predicted for higher beam velocities; see \citep{Shaaban_2018MNRAS, Shaaban_2018PoP} for exact thresholds derived numerically. 
The present analysis also invokes complex setups with intrinsic temperature ($T$) anisotropies, when HFIs interplay, for instance, with the whistler instability (WI) driven by $T_\perp > T_\parallel$ (where $\perp, \parallel$ indicate gyrotropic orientations with respect to the magnetic field $B_0$) or the firehose instability (FI) induced by an opposite anisotropy $T_\perp < T_\parallel$.
The new results with RKD distributed electrons are systematically presented and discussed in Section \ref{sec:results}. 
We thus prove the ability of ALPS to resolve pure WHFI and FHFI, as well as their interplay with temperature anisotropy instabilities such as WI and FI. 
Our results are summarized in the last section, Section~\ref{sec:summary}, which also provides a series of conclusions and prospects.


\section{Electron distributions and wave dispersion and stability formalism}\label{sec:theory}
The electron velocity distribution function (VDF) typically consists of a dense thermal core $f_c$ and a less dense drifting suprathermal component $f_s$ along the magnetic field. The total electron distribution function is modeled as
\begin{equation}\label{eq:core-halo model}
f_{}(v_{\parallel}, v_{\perp}) = \dfrac{n_c}{n} f_{c}(v_{\parallel}, v_{\perp}) +\dfrac{n_s}{n} f_{s}(v_{\parallel}, v_{\perp}).
\end{equation}
where \( n = n_c + n_s\) represents the total number density of plasma particles with $n_c$ and $n_s$ the densities of the core and strahl components, respectively, and $v_{\parallel,\perp}$ is the particle velocity parallel and perpendicular to the magnetic background field. In the proton rest frame, electrons form counterstreaming populations with drift velocities $u_c$ (core) and $u_s$ (strahl) with net zero current, which means $n_cu_c+n_su_s=0$, or $u_c=-\dfrac{n_s}{n_c}u_s$

The core is described by a drifting Maxwellian distribution $f_M$:
\begin{equation}
    f_c(v_\parallel, v_\perp) =f_M(v_\parallel, v_\perp)= \frac{1}{\pi^{3/2} \theta_{M,\perp}^2 \theta_{M,\parallel}} \exp\left(-\frac{(v_\parallel - u_c)^2}{\theta_{M,\parallel}^2} - \frac{v_\perp^2}{\theta_{M,\perp}^2}\right),
\end{equation}
where $\theta_{M \|, \perp} = \sqrt{2k_B T^M_{\|, \perp}/m}$ denotes the thermal speeds, with Boltzmann constant $k_B$, particle mass $m$, and the Maxwellian temperature $T^M_{\|, \perp}$. 

\begin{figure}[t!]
 \centering
\includegraphics[width=0.7\textwidth]{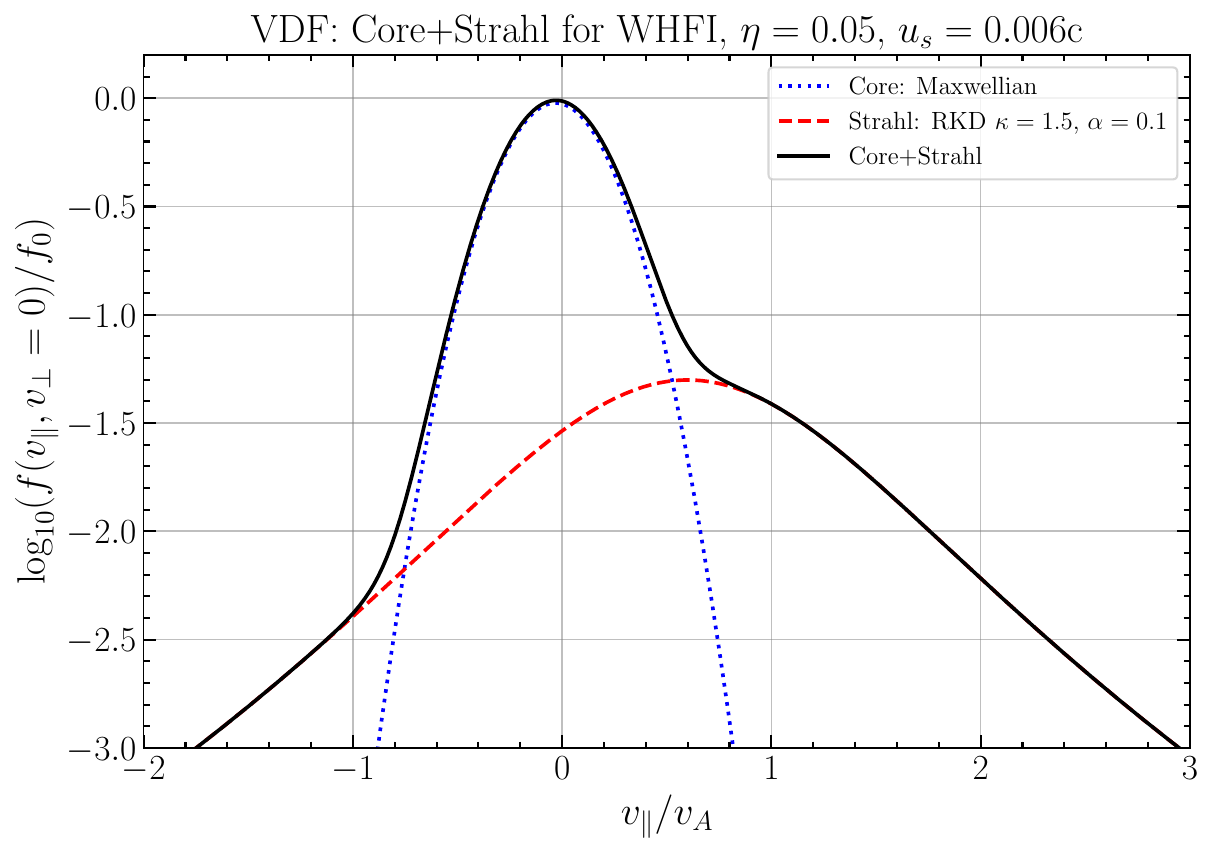}
\caption{\textit{Plot of the VDF Core-Strahl model. The Maxwellian core is plotted with blue dots, the RKD Strahl is dashed red, and the combination is shown in solid black.}
\label{fig:VDF_Core_Strahl_Model}}
\end{figure}

The strahl component follows a drifting Kappa distribution, which was previously widely modeled with the SKD \citep{Lazar_Fichtner_Yoon_2016}:
\begin{equation}\label{eq:fSKD}
     f_{\text{SKD}}(v_\parallel, v_\perp,\kappa) = \frac{1}{\pi^{3/2} \theta_\perp^2 \theta_\parallel} \frac{\Gamma(\kappa + 1)}{\Gamma(\kappa - 1/2)} \left[1 + \frac{(v_\parallel - u_s)^2}{\kappa \theta_\parallel^2} + \frac{v_\perp^2}{\kappa \theta_\perp^2}\right]^{-\kappa -1}.
\end{equation}
The parameter $\kappa$ is essential in determining the high-energy tails, and $\Gamma$ is the Gamma function.
The thermal speeds are defined as $\Theta_{\|,\perp}=\sqrt{2k_BT^{\kappa}_{\|,\perp}/m}$ with $T^{\kappa}_{\|,\perp}=\dfrac{\kappa}{\kappa-3/2}T^{M}_{\|,\perp}>T^{M}_{\|,\perp}$ \citep{Lazar-etal-2015}.
The magnetohydrodynamic equations in fluid theory can be derived from kinetic theory by computing velocity moments, which, even though there is an infinite number of them,  must exist for a valid physical description. While a Maxwellian distribution satisfies this requirement, an SKD only ensures convergence for moments of the order $l$ with $l<2\kappa -1$ \citep{Scherer2018}. This implies that for a well-defined kinetic temperature, which corresponds to the second moment $l=2$, the condition $\kappa > 3/2$ must hold. However, observations indicate that space plasmas can exhibit values of $\kappa \leq 3/2$ \citep{Gloeckler-etal-2012}. To address these limitations and provide a physically consistent framework, the RKD was introduced \citep{Scherer2018}. This more advanced model will, for the first time, also be used to describe the strahl component, as:

\begin{equation}\label{eq: frkd}
\begin{aligned}
f_{\mathrm{RKD}}\left(v_{\|}, v_{\perp},\kappa, \alpha \right)= &  \frac{1}{\pi^{3 / 2} \Theta_{\|} \Theta_{\perp}^2 W} \left(1+\frac{(v_{\|}-u_s)^2}{\kappa \Theta_{\|}^2}+\frac{v_{\perp}^2}{\kappa \Theta_{\perp}^2}\right)^{-\kappa-1} \exp \left(-\frac{\alpha_{}^2 (v_{\|}-u_s)^2}{\Theta_{\|}^2}-\frac{\alpha_{}^2 v_{\perp}^2}{\Theta_{\perp}^2}\right)
\end{aligned}
\end{equation}

with 
\begin{equation}
    W = U\left(\frac{3}{2}, \frac{3-2\kappa}{2},\alpha^2 \kappa \right)
\end{equation}
and $U$ representing the Tricomi function \citep{Scherer_2019}. The parameter $\alpha$ is a dimensionless {cutoff} parameter that controls the strength of the exponential decay and is independent of $\kappa$. The {Core-Strahl} model is plotted in Fig. (\ref{fig:VDF_Core_Strahl_Model}): The Maxwellian core (dotted blue), a less dense, hotter strahl component, represented with a counterstreaming RKD (dashed red), and the combined model (solid black).

The formalism for obtaining the dispersion relation in a general (numerical) form is explained in Appendix \ref{app_disper}, see also \cite{Verscharen-etal-2019, SchröderPoP2025}. 
The initial (unperturbed) distribution function (\ref{eq: frkd}) is introduced into plasma susceptibilities from Equation (\ref{Eq_sus}), and finally, the dispersion Equation~(\ref{A5}) is obtained. 
This is solved numerically to obtain the complex solutions
\begin{equation}
    \omega(\boldsymbol{k}) = \omega_r (\boldsymbol{k}) + i\gamma(\boldsymbol{k}).
\end{equation}
where $\boldsymbol{k}$ is the wavevector, $\omega_r=\Re(\omega)$ gives us the wave frequency, and $\gamma=\Im(\omega)$ the growth (or damping) rate of the instabilities (or damped waves).
In this work, we use ALPS to solve Equation~(\ref{A5}) numerically for quasi-parallel modes ($k_{\perp}\approx 0$). 
For more details on ALPS and its capability, refer to \cite{SchröderPoP2025} and \cite{Verscharen-etal-2019}.

\section{Validation}\label{sec:validation}

To validate the application of ALPS to heat-flux instabilities, we first use the DIS-K code \citep{Lopez_Shaaban_Lazar_2021, López2021_book, rodrigo_a_lopez_DISK}, developed to solve the dispersion tensor for drifting {SKD-distributed} plasmas explicitly, to reproduce the results in \cite{Shaaban_2018MNRAS, Shaaban_2018PoP}. 
The same results are then compared to those obtained with the ALPS code. 
This approach complements previous ALPS-based studies of temperature anisotropies in electron plasmas with {nondrifting} RKD halos \citep{SchröderPoP2025}.

The plasma frequency for the $j$th species is defined as $\omega_{p,j} = \sqrt{4 \pi n_j q_j^2 / m_j}$, while the nonrelativistic gyrofrequency is given by $\Omega_j = q_j B_0 / (m_j c)$. The reference Alfv\'{e}n speed is $v_{A,\text{ref}}/c = v_{A,j}/c = B_0 / (\sqrt{4\pi n_j m_j} c)$, where $m_j$ denotes the rest mass, $q_j$ the electric charge, and $n_j$ the number density of species $j$ ($j = p$ for protons and $j = e$ for electrons), and $c$ is the speed of light.
We employ parallel plasma betas defined as $\beta_{j,c} = 8 \pi n_c k_B T_{j,c,\|} / B_0^2$ and $\beta_{j,s} = 8 \pi n_s k_B T_{j,s,\|} / B_0^2$. Furthermore, we assume a proton-to-electron mass ratio of $m_p/m_e = 1836$, and set the electron plasma-to-gyrofrequency ratio at $\omega_{p,e}/|\Omega_e| = 100$. The thermal speed of the strahl electrons is expressed as $v_{\|,\text{th}}/c = \sqrt{\beta_s} |\Omega_e| / \omega_{p,e}$.
The core and strahl populations are assigned densities $n_c / n_0 = 0.95$ and $n_s / n_0 = 0.05$, respectively, where $n_0$ is the total electron density, {resulting in a contrast between the core and the strahl density of} $\eta = n_s / n_c \approx 0.05$.

For the electron velocity distribution function, we adopt the expression given in Equation~(\ref{eq:core-halo model}), with the strahl component represented by $f_s = f_{\text{SKD}}$ from Equation~(\ref{eq:fSKD}), constructing a consistent dual-Maxwellian-$\kappa$ model. Proton interactions are considered negligible and are thus modeled using an isotropic Maxwellian distribution. The parameters of the electron plasma used for the different cases in this study are listed in Table \ref{tab:parameters}.

\begin{deluxetable*}{lcccccc}
\tablecaption{Electron Plasma Parameters for the Different Cases.\label{tab:parameters}}
\tablehead{
\colhead{} & 
\colhead{$A_c$} & 
\colhead{$A_s$} & 
\colhead{$\beta_{e,c}$} & 
\colhead{$\beta_{e,s}$} & 
\colhead{$u_s/c$} & 
\colhead{$\eta$}
}
\startdata
WHFI & 1.0 & 1.0 & 0.1 & 0.05 & 0.006 & 0.05 \\
FHFI & 1.0 & 1.0 & 1.2 & 0.6 & 0.036 & 0.05 \\
WI   & 1.0 & 3.5 & 0.04 & 0.02 & 0.028 & 0.05 \\
EFHI & 1.0 & 0.2 & 4.0 & 2.0 & 0.005 & 0.05\\
\enddata
\end{deluxetable*}

\subsection{Whistler Heat-Flux Instability}\label{sec:validation_whfi}
With respect to Figure 1 in \cite{Shaaban_2018MNRAS}, we use $\beta_{ e, c}=\beta_p=0.1$, $\beta_{ e, s}=0.05$, and $T_{\perp, e, c}/T_{\|, e, c}=T_{\perp, e, s}/T_{\|, e, s}=1$ and a beaming velocity of $u_s=0.006c=2.68v_{\|,th}$. These parameters can be expected in the solar wind at $1$ au \citep{StverákJGP_2008}.

\begin{figure}[ht!]
 \centering
\includegraphics[width=0.9\textwidth]{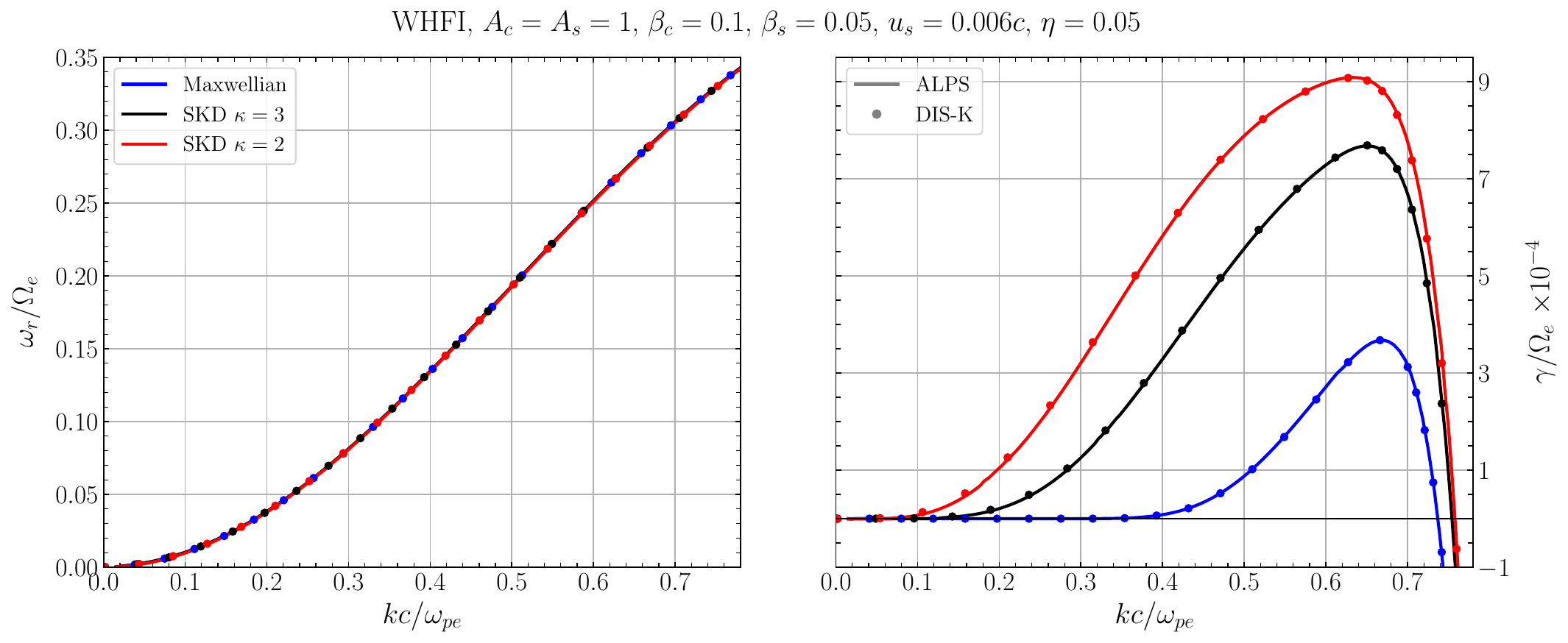}
\caption{\textit{Comparison of ALPS and DIS-K for WHFI with diﬀerent SKD strahl (blue: Maxwellian, black: SKD $\kappa=3$, red: SKD $\kappa=2$; solid:
results derived with ALPS, dots: DIS-K results). All parameters are stated above the panels.}
\label{fig:WHFI_Shaaban2018_b01_ub06_skd_vergleich}}
\end{figure}

The results are shown in Figure \ref{fig:WHFI_Shaaban2018_b01_ub06_skd_vergleich}, with frequencies on the left-hand side panel and the growth rates on the right-hand side panel. The DIS-K results are represented with dots, while solid lines are used for the ALPS results. The Maxwellian cases are plotted in blue, while the different SKD cases are plotted in black ($\kappa=3$) and red. The agreement for both the frequency and the growth rate is excellent for all cases, capturing the fastest growing mode and showing the correct analytical continuation. This serves to validate not only the ALPS for heat-flux studies but also the implementation of our core–strahl model, thereby reinforcing the RKD-based results presented in the following Section. 

\subsection{Electron Firehose Heat-Flux Instability}\label{sec:validation_fhfi}
\begin{figure}[ht!]
 \centering
\includegraphics[width=0.9\textwidth]{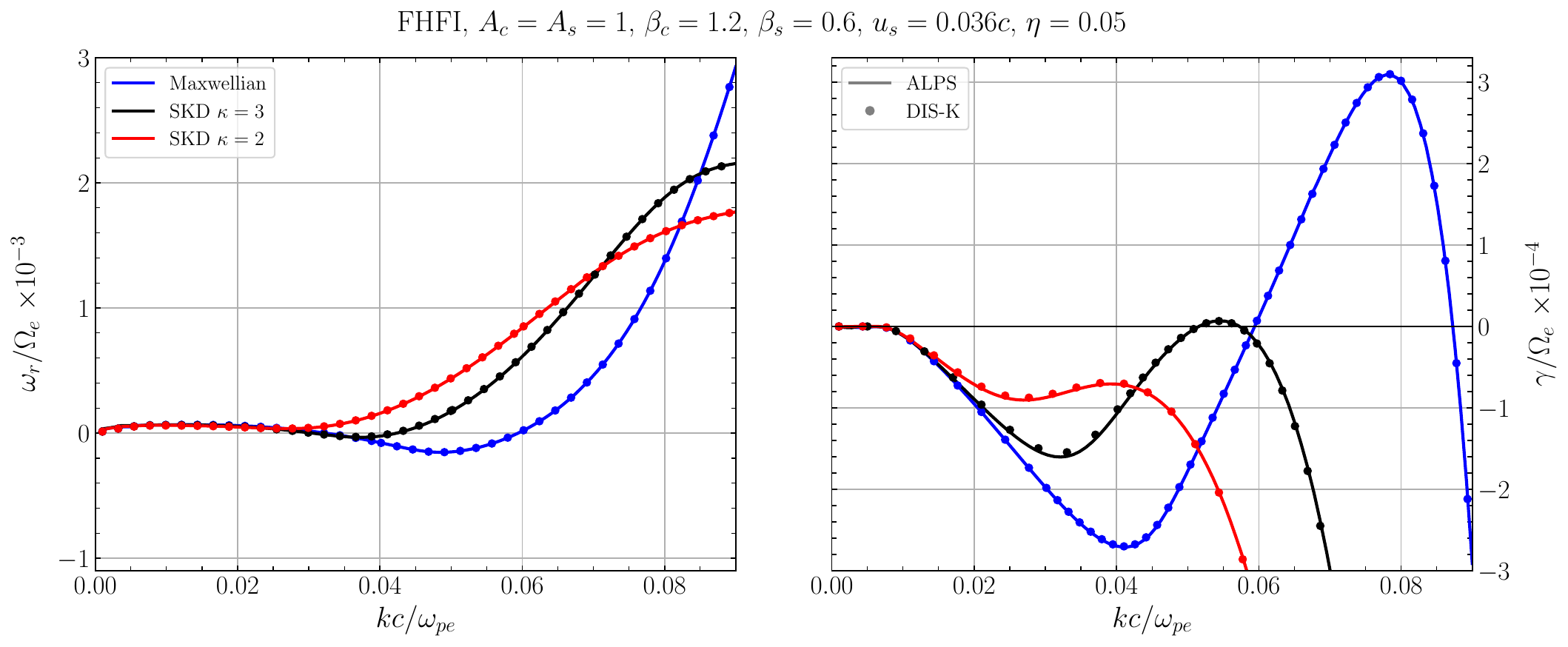}
\caption{\textit{Comparison of ALPS and DIS-K for FHFI with different SKD strahl (blue: Maxwellian, black: SKD $\kappa=3$, red: SKD $\kappa=2$; solid:
results derived with ALPS, dots: DIS-K results). All parameters are stated above the panels.} 
\label{fig:FHFI_Shaaban2018_b12_ub36_skd_vergleich}}
\end{figure}

With reference to Figure 8 in \cite{Shaaban_2018MNRAS}, we use $\beta_{ e, c}=1.2$, $\beta_{ e, s}=0.6$, and $T_{\perp, e, c}/T_{\|, e, c}=T_{\perp, e, s}/T_{\|, e, s}=1$, and a beam velocity of $u_s=0.036c=4.65v_{\|,th}$. 
The results are shown in Figure \ref{fig:FHFI_Shaaban2018_b12_ub36_skd_vergleich}, with frequencies on the left-hand side panel and the growth rates on the right-hand side panel. Again, the DIS-K results are represented with dots, while solid lines are used for the ALPS results. The Maxwellian cases are plotted in blue, while the different SKD cases are plotted in black and red. The agreement is very good.

\subsection{Temperature anisotropy + Heat flux}\label{sec:validation_temp_aniso}
While the whistler mode arises for a temperature excess in the direction perpendicular to the magnetic field, $A = T_\perp/T_\parallel >1$, at lower beam velocities, the firehose instability exists for the opposite anisotropy $A<1$.
\subsubsection{Whistler instability}
\begin{figure}[ht!]
 \centering
\includegraphics[width=0.9\textwidth]{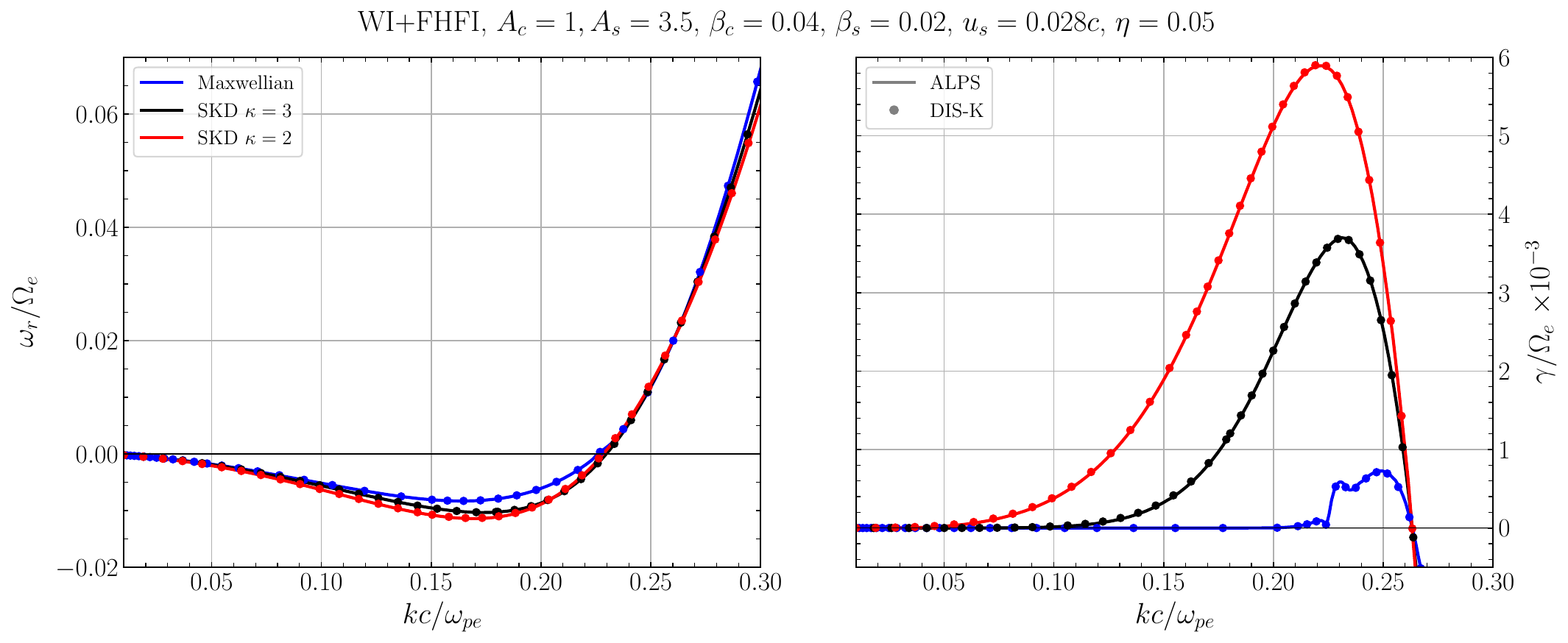}
\caption{\textit{Comparison of ALPS and DIS-K for WI+heat flux with different SKD strahl (blue: Maxwellian, black: SKD $\kappa=3$, red: SKD $\kappa=2$ ; solid:
results derived with ALPS, dots: Dis-K results). All parameters are stated above the panels.}
\label{fig:FHFI_WI_Shaaban2018_Ah35_b004_ub28_skd_vergleich}}
\end{figure}

For WI, we use $\beta_{ e, c}=0.04$, $\beta_{ e, s}=0.02$, and $T_{\perp, e, c}/T_{\|, e, c}=1$, $T_{\perp, e, s}/T_{\|, e, s}=3.5$, and a strahl velocity of $u_s=0.028c=19.8v_{\|,th}$. This represents Figure 7 in \cite{Shaaban_2018PoP}. The results are shown in Figure \ref{fig:FHFI_WI_Shaaban2018_Ah35_b004_ub28_skd_vergleich}, with the same format as in the previous figures. Once again, the agreement between the solvers is excellent.

\begin{figure}[ht!]
 \centering
\includegraphics[width=0.9\textwidth]{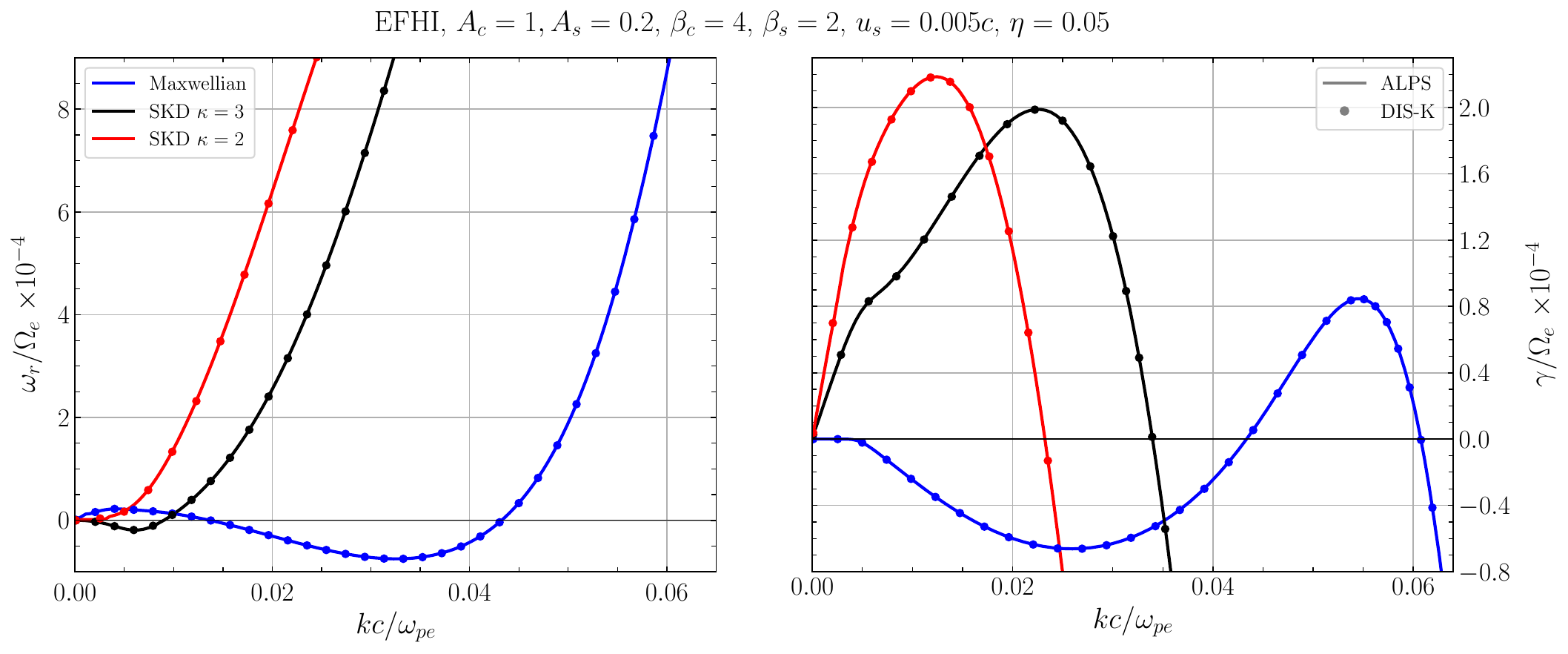}
\caption{\textit{Comparison of ALPS and DIS-K for EFHI with different SKD strahl (blue: Maxwellian, black: SKD $\kappa=3$, red: SKD $\kappa=2$ ; solid:
results derived with ALPS, dots: DIS-K results). All parameters are stated above the panels.}
\label{fig:EFHI_Shaaban2018_Ab02_b4_ub0005_skd_vergleich}}
\end{figure}

\subsubsection{Electron firehose instability}

For EFHI, referring to  Figure 8 in \cite{Shaaban_2018PoP}, we use $\beta_{ e, c}=4$, $\beta_{ e, s}=2$, and $T_{\perp, e, c}/T_{\|, e, c}=1$, $T_{\perp, e, s}/T_{\|, e, s}=0.2$, and a strahl velocity of $u_s=0.005c=2.55v_{\|,th}$. The results are shown in Figure \ref{fig:EFHI_Shaaban2018_Ab02_b4_ub0005_skd_vergleich}, with the same format as before, providing a very good agreement.

Notice that the agreement between solvers is not only for the unstable part of the solutions ($\gamma>0$), but also for the damped part ($\gamma<0$). That also shows that the analytical continuation of both solvers is working correctly, which is a crucial feature in this kind of dispersion solver.

\section{UNSTABLE SOLUTIONS WITH DRIFTING REGULARIZED KAPPA DISTRIBUTION  Strahl}\label{sec:results}
Now that we have shown the validity of the ALPS code to model Kappa distributed plasmas, we will extend our analysis to include the effect of RKD electrons.
We model the electron strahl population with a drifting regularized bi-$\kappa$-distribution, as defined in Equation~\eqref{eq: frkd}. The RKD cases provide an opportunity to delve deeper into the impact of modifying the suprathermal tail through the parameters $\alpha$ and $\kappa$, showing their interplay in determining plasma (in)stability. The parameters for the different cases are the same as for the SKD calculations, see Table \ref{tab:parameters}. RKDs with the same $\alpha$ are plotted in the same dashed style, while the Maxwellian and Maxwellian-like curves are dashed-dotted.
\subsection{Whistler Heat-Flux Instability}
\begin{figure}[ht!]
 \centering
\includegraphics[width=0.9\textwidth]{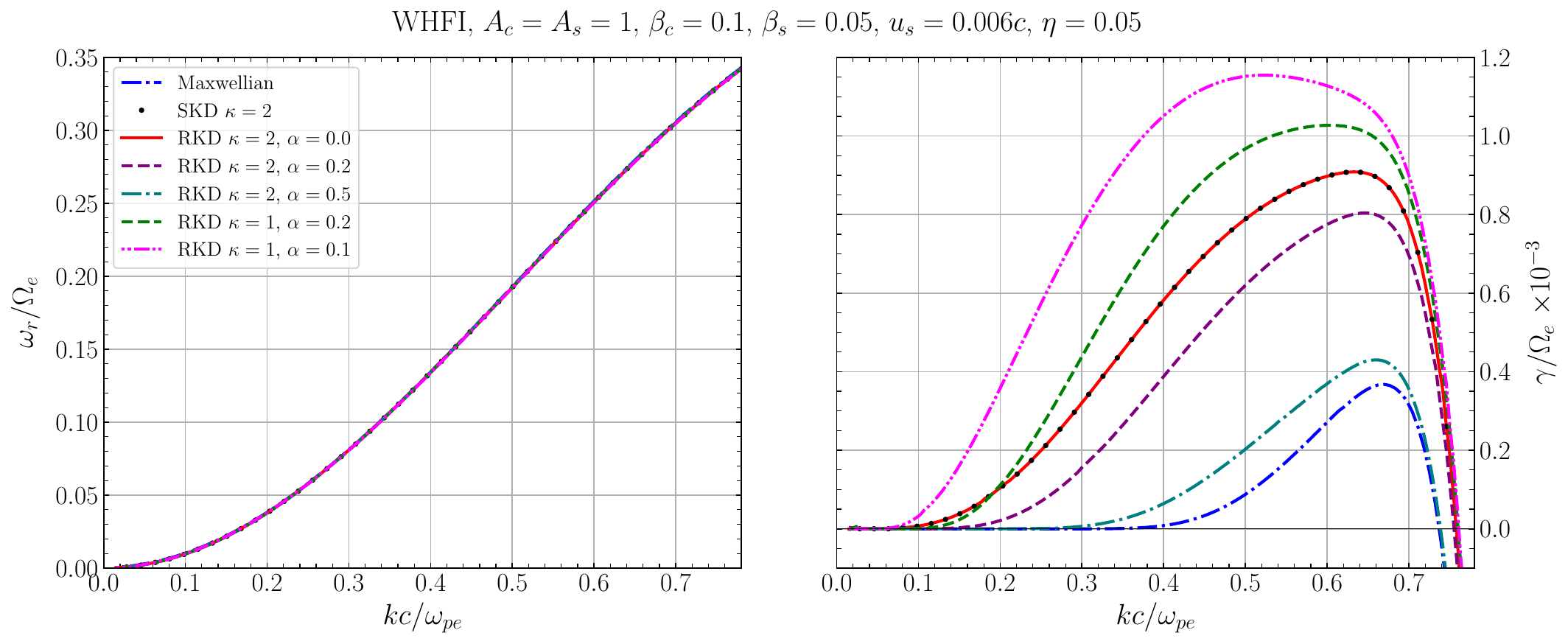}
\caption{\textit{WHFI with different RKD strahl with reference to Figure \ref{fig:WHFI_Shaaban2018_b01_ub06_skd_vergleich}. RKDs with the same $\alpha$ are represented using the same dashed style, whereas the Maxwellian and Maxwellian-like RKD are shown with a dashed-dotted line. The RKD without a cutoff (dotted) produces the same result as the SKD and converges toward the Maxwellian solution as the cutoff increases, demonstrating a distinct ordering in $\alpha$ concerning the maximum growth rate. Increasing $\kappa$  amplifies the growth rates noticeably.}
\label{fig:WHFI_Shaaban2018_b01_ub06_rkd}}
\end{figure}

\begin{figure}[ht!]
 \centering
\includegraphics[width=0.5\textwidth]{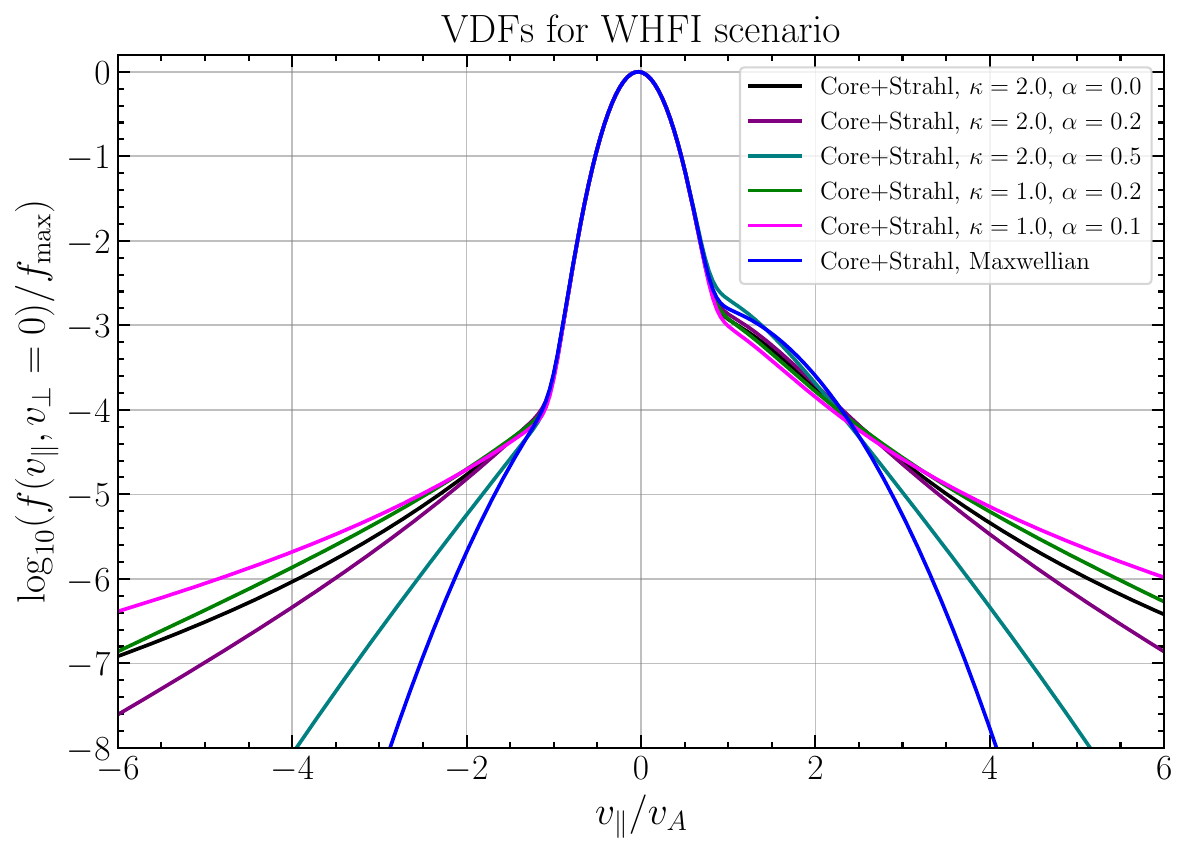}
\caption{\textit{Plot of the different VDFs with a counterstreaming strahl normalized to their maximum for the WHFI case corresponding to Figure \ref{fig:WHFI_Shaaban2018_b01_ub06_rkd}.}
\label{fig:VDF_WHFI}}
\end{figure}

\begin{figure}[ht!]
 \centering
\includegraphics[width=1.0\textwidth]{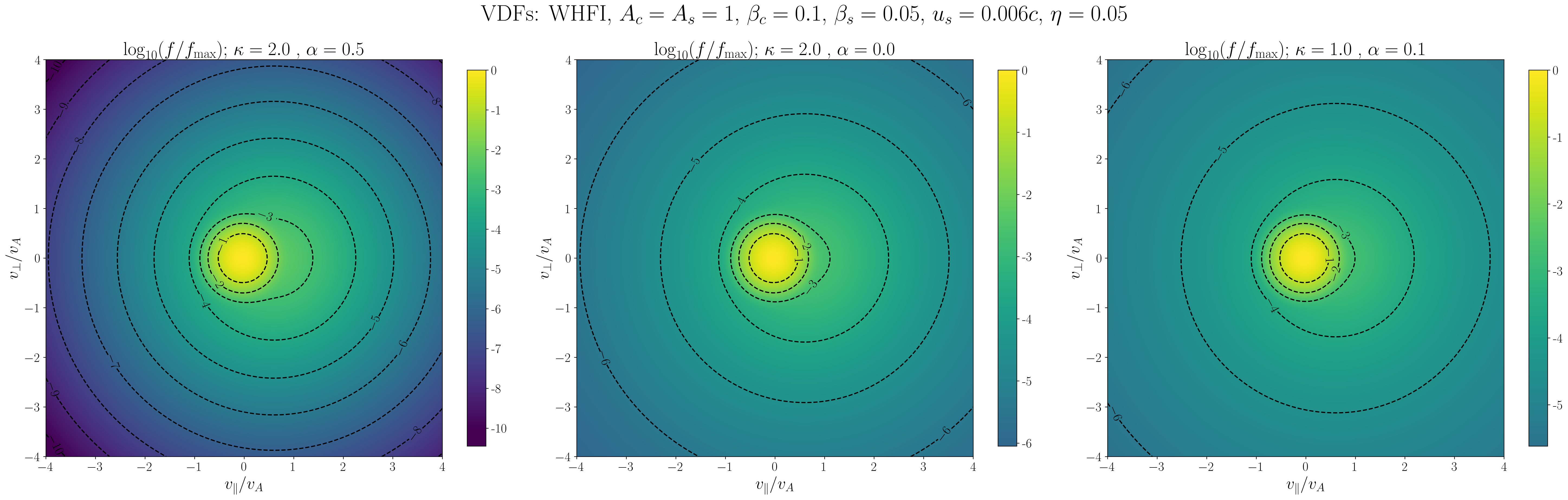}
\caption{\textit{Contour plots of three different RKD core-strahl VDFs normalized to their maximum for the WHFI case. With $((\kappa=2,\alpha=0.5);(\kappa=2, \alpha=0.0);(\kappa=1, \alpha=0.1))$ from left to right. The anisotropy generated through the counterstreaming strahl is clearly visible. Note
that actually $v_{\perp}>0$ due to the cylindrical coordinate system used in ALPS, but for illustrative purposes, values  $v_{\perp}<0$ are also shown.}
\label{fig:VDF_contour_WHFI}}
\end{figure}

The results for the WHFI resolved using ALPS are shown in Figure \ref{fig:WHFI_Shaaban2018_b01_ub06_rkd}, with frequencies on the left-hand side panel and the growth rate on the right-hand side panel. The RKD with $\kappa =2$ and no cutoff $\alpha =0$ (black dots) leads to a very good agreement, for both frequency and growth rate, with the results of the SKD with $\kappa=2$ (red solid curve), validating the implementation of the RKD model, as expected.\\
As shown in \cite{Shaaban_2018MNRAS}, the choice of the suprathermal strahl component should strongly impact the growth rate. 
The RKDs display distinct variations in growth rates depending on the parameters $\kappa$ and $\alpha$. Increasing $\alpha$ at fixed $\kappa$ and thus rendering the distribution increasingly Maxwellian-like, systematically reduces the overall growth rates. This manifests as a lower maximum growth rate, shifted toward higher wavenumbers, reflecting the suppression of instabilities due to diminished high-energy tails and thus a less stronger counter stream.\\
As anticipated and shown in \cite{SchröderPoP2025}, with larger $\alpha$, the RKD progressively converges toward Maxwellian behavior, establishing a clear hierarchy in maximum growth rates with respect to the cutoff parameter $\alpha$.
Conversely, smaller values of \(\alpha\), when combined with lower \(\kappa\) (e.g., \(\kappa = 1.0\)), enhance the nonthermal character of the distribution, resulting in substantially higher growth rates. 
For \(\kappa = 1.0\) and \(\alpha = 0.2\) (green dashed curve), the dominance of the high-energy tail is sufficient to produce substantially higher growth rates, even in the presence of a moderate {cutoff}. This trend is amplified for \(\kappa = 1.0\) and \(\alpha = 0.1\) (magenta double-dotted dashed curve), where the growth rate develops into a broader and more elevated plateau-like peak, revealing enhanced instability across an extended range of wavenumbers, even so for small wavenumbers. Notably, the high growth rates observed in these RKD configurations would not be attainable with an SKD.\\
The various core-strahl VDFs, given by $f_{} = ({n_c}/{n}) f_{c}+({n_s}/{n}) f_{s}$, normalized to their respective maximum, are displayed in Figure \ref{fig:VDF_WHFI}. Corresponding contour plots for three selected VDFs, $(\kappa=2,\alpha=0.5);(\kappa=2, \alpha=0.0);(\kappa=1, \alpha=0.1)$ from left to right, are shown in Figure \ref{fig:VDF_contour_WHFI}. The anisotropy induced through the counterstreaming strahl is clearly visible in both Figure \ref{fig:VDF_WHFI} and \ref{fig:VDF_contour_WHFI}. Additionally, the influence of lower $\kappa$-values and higher $\alpha$ is evident: the decrease of $f$ with increasing velocity $v_{\|}$ is noticeably less steep for VDFs with $\kappa=1.0$ (green and magenta curves) compared to $\kappa= 2.0.$ (purple curve), while the VDF with $\alpha=0.5$ (teal curve) shows a much sharper decrease of the density with increasing $v_{\|}$, comparable with the Maxwellian case (blue {curve}). This is in expected agreement with the obtained results, i.e., VDFs with a higher density at high velocities lead to higher growth rates.        

\subsection{Firehose Heat-Flux Instability}
\begin{figure}[ht!]
 \centering
\includegraphics[width=0.9\textwidth]{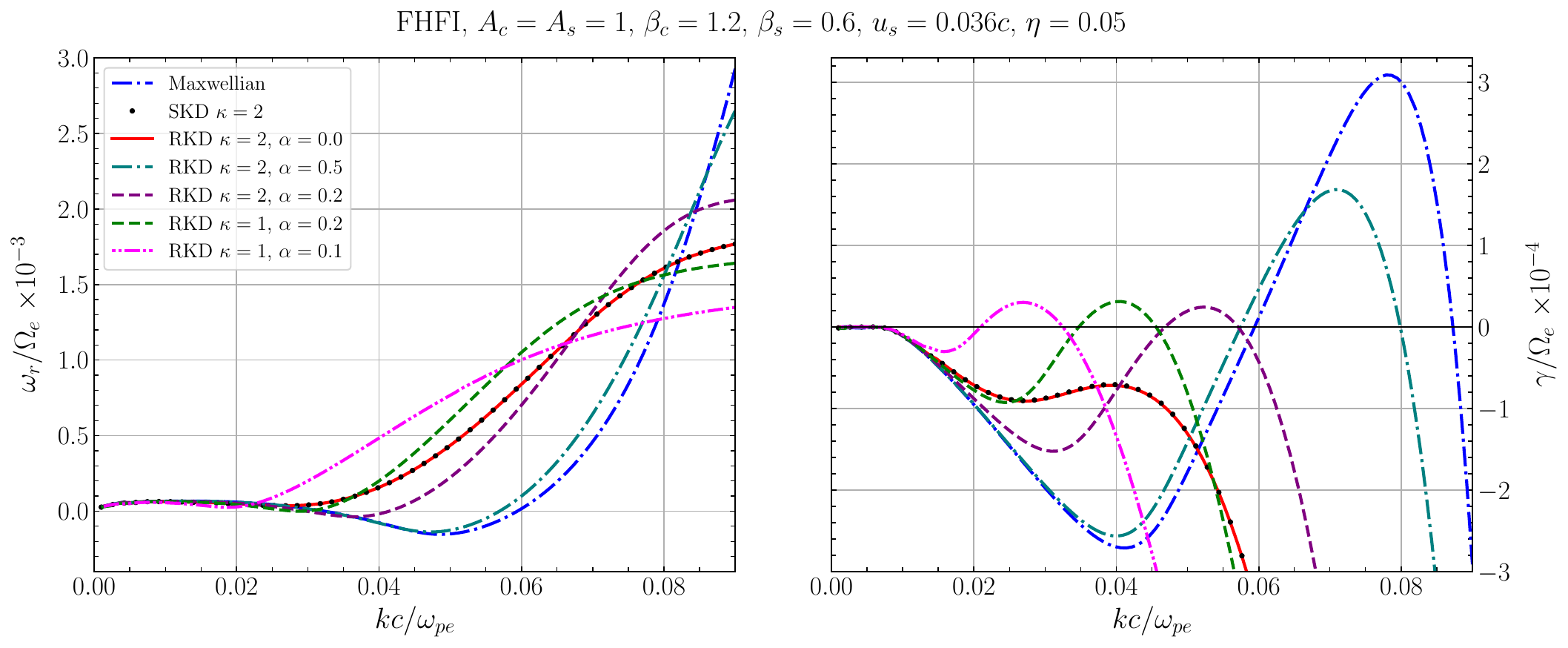}
\caption{\textit{FHFI with different RKD strahl with reference to Figure \ref{fig:FHFI_Shaaban2018_b12_ub36_skd_vergleich}. RKDs with the same $\alpha$ are represented using the same dashed style, whereas the Maxwellian and Maxwellian-like RKD are shown with a dashed-dotted line. The RKD without a cutoff (dotted) produces the same result as the SKD and converges toward the Maxwellian solution as the cutoff increases, demonstrating a distinct ordering in $\alpha$ concerning the maximum growth rate. Increasing $\kappa$ while having a nonzero value for $\alpha$ leads to unstable solutions, in contrast to the SKD cases.}
\label{fig:FHFI_b12_ub36_rkd}}
\end{figure}

\begin{figure}[ht!]
 \centering \includegraphics[width=0.5\textwidth]{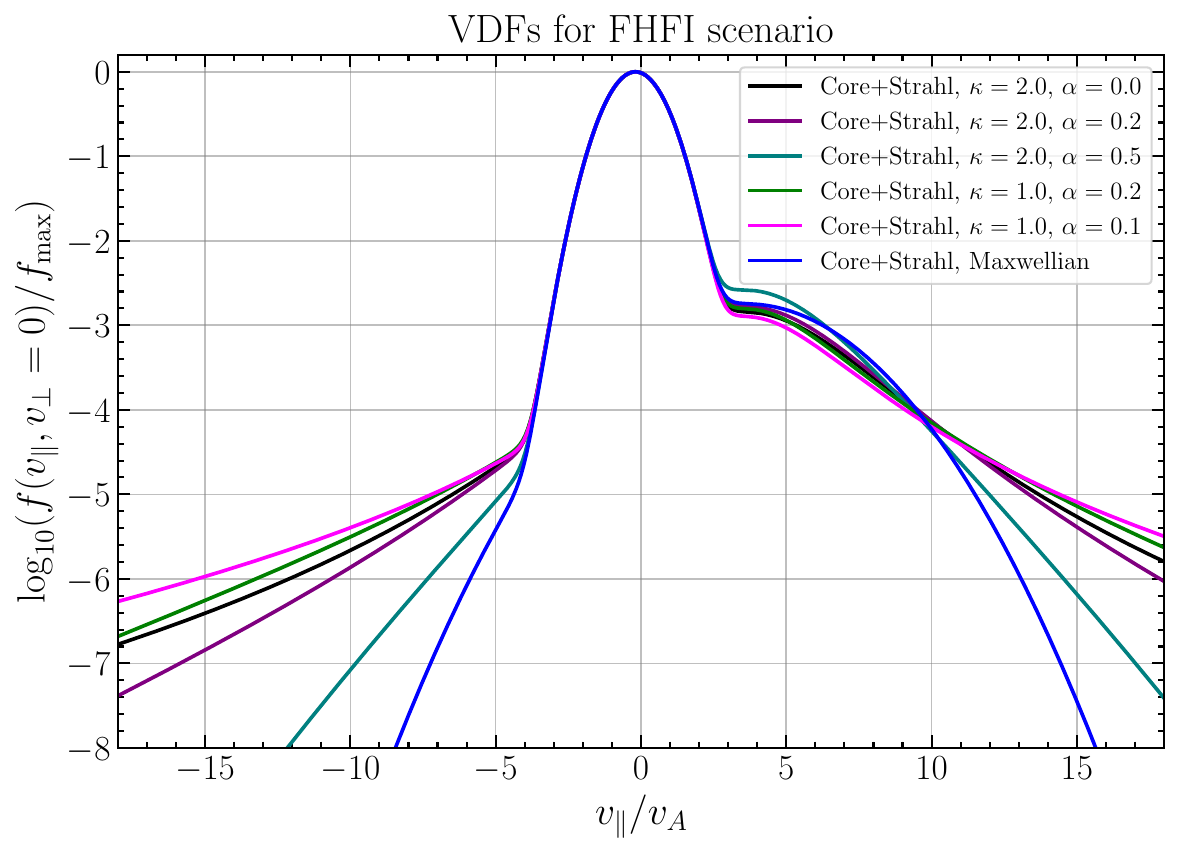}
\caption{\textit{Plot of the different VDFs with a counterstreaming strahl normalized to their maximum for the FHFI case corresponding to Figure \ref{fig:FHFI_b12_ub36_rkd}.}
\label{fig:VDF_FHFI}}
\end{figure}

\begin{figure}[ht!]
 \centering
\includegraphics[width=1.0\textwidth]{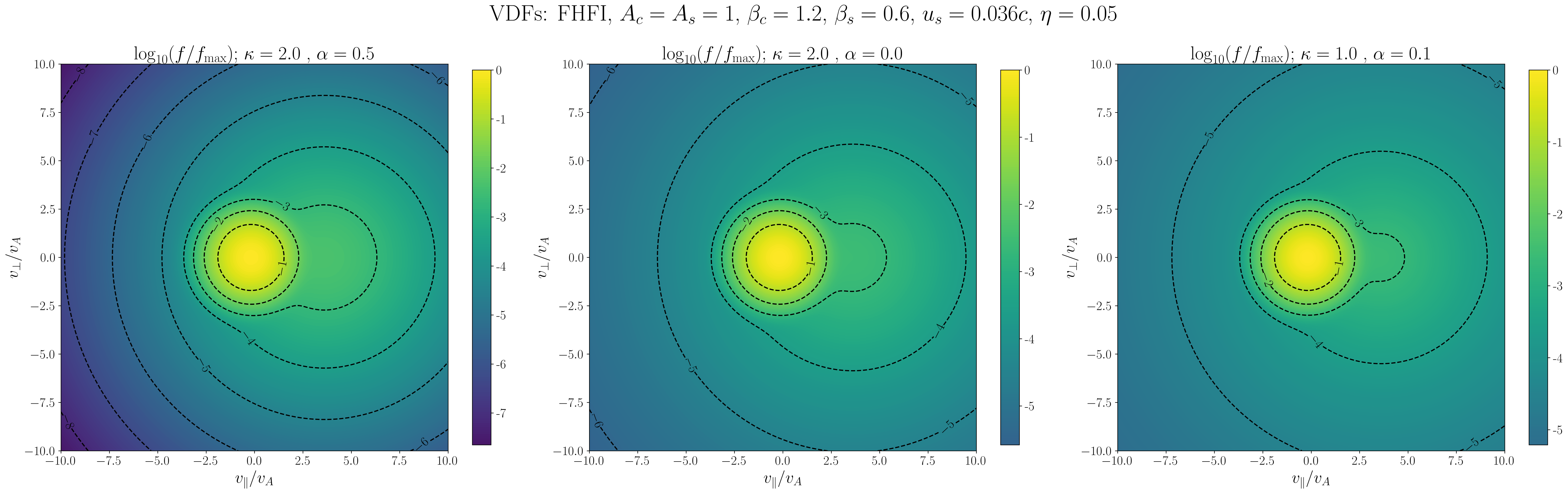}
\caption{\textit{Contour plots of three different RKD core-strahl VDFs normalized to their maximum for the FHFI case. With $((\kappa=2,\alpha=0.5);(\kappa=2, \alpha=0.0);(\kappa=1, \alpha=0.1))$ from left to right. The anisotropy generated through the counterstreaming strahl is clearly visible.  Note
that actually $v_{\perp}>0$ due to the cylindrical coordinate system used in ALPS, but for illustrative purposes, values  $v_{\perp}<0$ are also shown.}
\label{fig:VDF_contour_FHFI}}
\end{figure}

The results are presented in Figure \ref{fig:FHFI_b12_ub36_rkd}, with frequency on the left-hand side panel and the growth rate on the right-hand side panel. A very good agreement is achieved between the results for the SKD with $\kappa=2$ (red solid curve) and the RKD with $\kappa=2$ and $\alpha =0.0$ (black dots). Cutting into the suprathermal strahl by setting $\alpha =0.2$ (purple dashed curve), the evolution of both $\omega$ and $\gamma$ becomes comparable to that of the SKD with $\kappa = 3$, though the growth rate is slightly enhanced. A stronger {cutoff}, $\alpha = 0.5$ (teal dashed-dotted curve), results in a behavior similar to the Maxwellian results, in both real frequency and growth rate. The latter exhibits a clearly elevated maximum, occurring at higher wavenumbers. Employing a more suprathermal RKD with $\kappa = 1$ and $\alpha = 0.2$ (green dashed curve) develops, in contrast to the RKD with $\kappa =2$ without a {cutoff}, unstable modes with a maximum growth rate that is comparable to, but slightly higher than, that of the RKD with $\kappa = 2$ and $\alpha = 0.2$, and shifted toward lower wavenumbers. Compared to the RKD cases with $\kappa = 2$, the real frequency increases more steeply but saturates at lower wavenumbers. Lowering the {cutoff} further to $\alpha = 0.1$ (magenta double-dotted dashed curve) also results in unstable modes; however, the maximum growth rate is not significantly altered, only shifted toward lower values of $k$. The real frequency increases sharply at first, but quickly transitions into a plateau.\\
These results indicate that the SKD model underestimates the influence of the FHFI, predicting only stable modes for low values of $\kappa$. In contrast, the RKD allows for the development of unstable modes even in the presence of a stronger suprathermal component, such as for $\kappa = 1$ with a moderate {cutoff}.\\
The different core-strahl VDFs $f_{} = ({n_c}/{n}) f_{c}+({n_s}/{n}) f_{s}$, normalized to their maximum, can be seen in Figure \ref{fig:VDF_FHFI} and in Figure \ref{fig:VDF_contour_FHFI} the contour plot for three of these VFDs, $(\kappa=2,\alpha=0.5);(\kappa=2, \alpha=0.0);(\kappa=1, \alpha=0.1)$ from left to right, are shown. 

\subsection{Temperature Anisotropy + Heat Flux}
\begin{figure}[ht!]
 \centering
\includegraphics[width=0.9\textwidth]{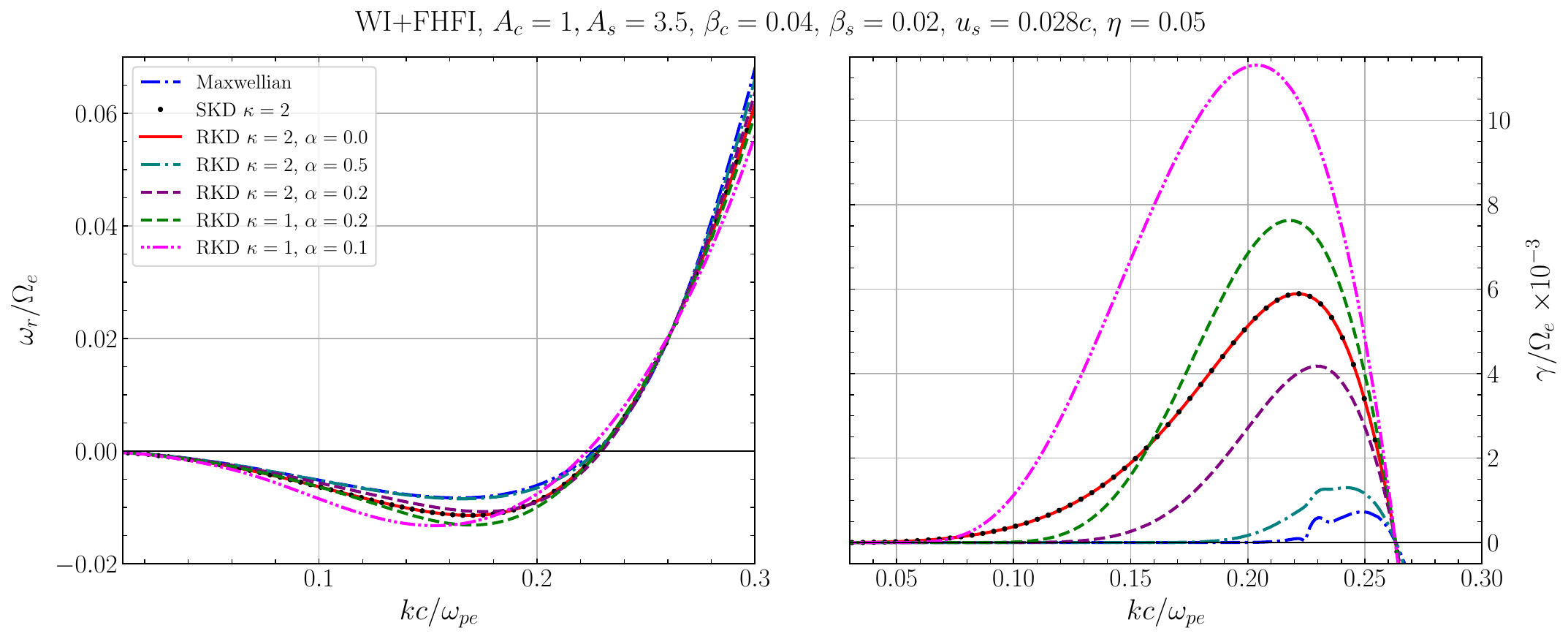}
\caption{\textit{WI + Heat flux with different RKD strahl with reference to Figure \ref{fig:FHFI_WI_Shaaban2018_Ah35_b004_ub28_skd_vergleich}. RKDs with the same $\alpha$ are represented using the same dashed style, whereas the Maxwellian and Maxwellian-like RKD are shown with a dashed-dotted line. The RKD without a cutoff (dotted) produces the same result as the SKD and converges toward the Maxwellian solution as the cutoff increases, demonstrating a distinct ordering in $\alpha$ concerning the maximum growth rate. Increasing $\kappa$ amplifies the growth rates noticeably.}
\label{fig:FHFI_WI_Ah35_b004_ub28_rkd}}
\end{figure}

\begin{figure}[ht!]
 \centering
\includegraphics[width=0.9\textwidth]{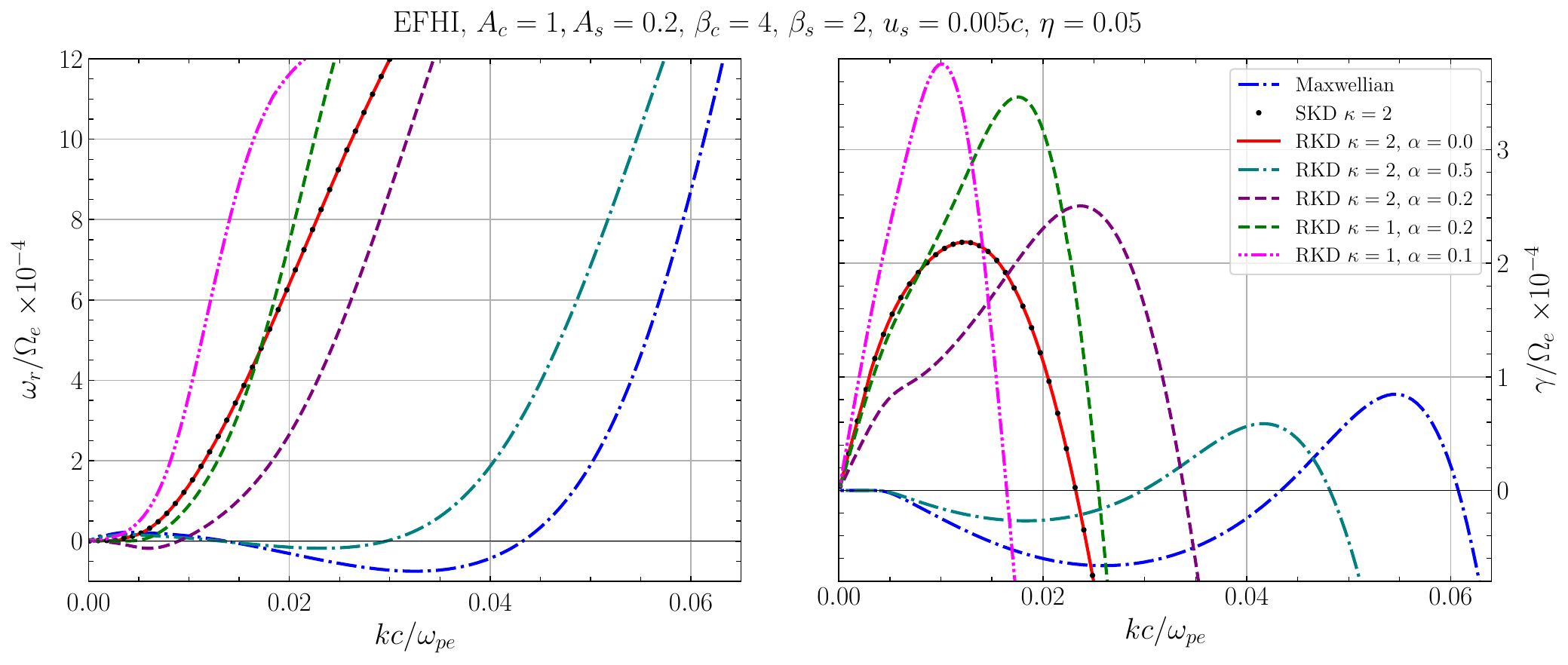}
\caption{\textit{EHFI with different RKD strahl with reference to Figure \ref{fig:EFHI_Shaaban2018_Ab02_b4_ub0005_skd_vergleich}. RKDs with the same $\alpha$ are represented using the same dashed style, whereas the Maxwellian and Maxwellian-like RKD are shown with a dashed-dotted line. The RKD without a cutoff (dotted) produces the same result as the SKD and converges toward the Maxwellian solution as the {cutoff} increases, demonstrating a distinct ordering in $\alpha$ concerning the maximum growth rate. Increasing $\kappa$  clearly amplifies the growth rates.}
\label{fig:EFHI_Ab02_b4_ub0005_rkd}}
\end{figure}

The results for WI are shown in Figure \ref{fig:FHFI_WI_Ah35_b004_ub28_rkd}, with frequency on the left-hand side panel and the growth rate on the right-hand side panel, while the results for EFHI are shown in Figure \ref{fig:EFHI_Ab02_b4_ub0005_rkd}. 
\subsubsection{Whistler Instability}
The results for the RKD with $\kappa=2$ and $\alpha =0$ (black dots) are in very good agreement with the SDK results (red solid curve). When the cutoff is set to $\alpha=0.2$, the overall growth rate is reduced and the maximum growth rate is shifted toward higher wavenumbers. When the cutoff is further increased to  $\alpha=0.5$, the growth rate is significantly suppressed: $\gamma$ becomes comparable to that of the Maxwellian case. Due to the diminished influence of the high-energy tail, the peak associated with the FHFI mode emerges more distinctly. When the suprathermal electron population is enhanced, as in the case of $\kappa = 1.0$ with $\alpha = 0.2$ (green dashed curve), the maximum growth rate increases and shifts toward lower wavenumbers. However, due to the presence of a moderate {cutoff}, the growth rate at the lowest wavenumbers drops off more steeply, falling below that of the RKD with $\kappa = 2$ and no {cutoff}. Once the {cutoff} is reduced to $\alpha = 0.1$ (magenta double-dotted dashed curve), the overall growth rate increases significantly, with the maximum growth rate becoming higher and shifting further toward smaller wavenumbers. However, the range of wavenumbers for which $\gamma > 0$ remains largely comparable to that of the RKD with $\kappa = 2$ and no {cutoff}. The real frequency exhibits only minor variations across the different cases. 
These results underscore the sensitivity of the whistler instability to the shape of the suprathermal electron distribution, demonstrating that lower values of $\kappa$ and $\alpha$ can substantially amplify the growth rate. 

\subsubsection{Electron Firehose Instability}
As expected, in the absence of a cutoff, the RKD with $\kappa = 2$ (black dots) yields results that closely match those of the SKD with $\kappa = 2$ (red solid curve). When the suprathermal population is reduced by introducing a {cutoff} of $\alpha = 0.2$ (purple dashed curve), the real frequency and growth rate resemble those of the SKD with $\kappa = 3$, albeit with a higher maximum growth rate. Using $\alpha = 0.5$ yields results for both $\omega$ and $\gamma$ that are similar to those of the Maxwellian case, characterized by a reduced growth rate and a shift toward higher wavenumbers. Notably, the maximum growth rate is even lower than in the Maxwellian case, likely because the high {cutoff} value $\alpha = 0.5$ suppresses part of the anisotropic population, thereby diminishing its destabilizing effect. When a lower $\kappa$ value ($\kappa = 1$) is used in combination with a {cutoff} of $\alpha = 0.2$ (green dashed curve), the growth rate develops a more sharply defined peak, accompanied by an increased maximum value. The real frequency rises at higher wavenumbers, but exhibits a steeper increase compared to the RKD case with $\alpha = 0$. The range of wavenumbers for which $\gamma>0$ is comparable to that for $\alpha=0.2$. An even weaker {cutoff}, $\alpha = 0.1$ (magenta double-dotted dashed curve), leads to pronounced changes: the growth rate exhibits a sharp peak with a higher maximum value, shifted toward lower wavenumbers. The range of $k$ for which $\gamma > 0$ is reduced, while the real frequency increases steeply at low wavenumbers.

\section{Summary}\label{sec:summary}
In this study, we investigated the linear dispersion properties of parallel heat-flux instabilities, specifically the electron firehose heat-flux instability and the whistler heat-flux instability, in a plasma composed of a core-strahl electron population modeled using a regularized $\kappa$-distribution. By varying the RKD parameter $\kappa$ and the {cutoff} $\alpha$, we systematically show how suprathermal features influence both the real frequency and growth rate of these instabilities.

Our results reveal that the RKD converges toward SKD or Maxwellian-like behaviour for larger $\kappa$ and higher $\alpha$, and can amplify growth rates for lower $\kappa$ and $\alpha$ values, providing a significantly more flexible framework for modeling electron velocity distributions in space plasmas. 
All cases for $\kappa < 3/2$ would not be accessible with an SKD model. Notably, we demonstrate that for FHFI for $\kappa < 3/2$, unstable modes can still develop in the RKD framework, whereas such {behavior} is absent in SKD models. This highlights the RKD’s capability to capture a broader range of {nonthermal} effects.

These findings further validate the ALPS solver as a robust and powerful tool for plasma instability studies. In particular, this work extends ALPS’ application to more cases, confirming its utility for investigating kinetic instabilities. Future work will extend the present analysis by exploring and comparing instabilities for RKD models with an anisotropic {cutoff} parameter. Additionally, a more comprehensive three-component model, incorporating a core, halo, and strahl, will be investigated to better represent observed electron distributions in the solar wind. Such an approach could yield deeper insights into the interplay between multiple {nonthermal} populations and the resulting kinetic instabilities.

\section*{ACKNOWLEDGMENTS}
The authors acknowledge support from the Ruhr-University Bochum, the Katholieke Universiteit Leuven, and the use of the ALPS code. This project was funded by the Deutsche Forschungsgesellschaft (DFG), project FI $706/31$-$1$, and the Belgian FWO-Vlaanderen G$002523$N, and SIDC Data Exploitation (ESA Prodex), No. $4000145223$. R.A.L. thanks the support of ANID, Chile, through Fondecyt grant No. 1251712, and by the International Space Science Institute (ISSI) in Bern, through ISSI International Team project 'Excitation and Dissipation of Kinetic-Scale Fluctuations in Space Plasmas' (ISSI Team project \#$24$-$612$). The authors gratefully acknowledge Daniel Verscharen and Kris Klein for their continuous support with ALPS, particularly for valuable discussions, assistance with technical questions, and ongoing updates to their solver. The authors also express their appreciation to Rudi Gaelzer for his extensive work on deriving formulae related to RKDs and ongoing discussions, which contributed to the depth of this research. {The authors thank the reviewer for providing a fast and constructive review.}\\
\textit{Software}: ALPS \citep{ALPS:2023}, DIS-K \citep{rodrigo_a_lopez_DISK}.
          

\appendix


\section*{Dispersion relation in ALPS} \label{app_disper}
The susceptibilities may be written as \citep{Verscharen_2018}
\begin{widetext}
\begin{equation}\label{Eq_sus}
\boldsymbol{\chi}_j=\frac{\omega_{\mathrm{p} j}^2}{\omega \Omega_{j}} \int_0^{\infty} 2 \pi p_{\perp} \mathrm{d} p_{\perp} \int_{-\infty}^{+\infty} \mathrm{d} p_{\|}\left[\hat{\boldsymbol{e}}_{\|} \hat{\boldsymbol{e}}_{\|} \frac{\Omega_{j}}{\omega}\left(\frac{1}{p_{\|}} \frac{\partial f_{0 j}}{\partial p_{\|}}-\frac{1}{p_{\perp}} \frac{\partial f_{0 j}}{\partial p_{\perp}}\right) p_{\|}^2\right. \\
\left.+\sum_{n=-\infty}^{+\infty} \frac{\Omega_{j} p_{\perp} U}{\omega-k_{\|} v_{\|}-n \Omega_{j}} \boldsymbol{T}_n\right].
\end{equation}
\end{widetext}
Here, $f_{0j}$ represents the background distribution function of a species $j$, $p_{\|,\perp}$ is the parallel and perpendicular momentum, and $\omega=\omega(\boldsymbol{k})$. The tensor $\boldsymbol{T}_n$ is defined as
$$
\boldsymbol{T}_n \equiv\left(\begin{array}{ccc}
\frac{n^2 J_n^2}{z^2} & \frac{i n J_n J_n^{\prime}}{z} & \frac{n J_n^2 p_{\|}}{z p_{\perp}} \\
-\frac{i n J_n J_n^{\prime}}{z} & \left(J_n^{\prime}\right)^2 & -\frac{i J_n J_n^{\prime} p_{\|}}{p_{\perp}} \\
\frac{n J_n^2 p_{\|}}{z p_{\perp}} & \frac{i J_n J_n^{\prime} p_{\|}}{p_{\perp}} & \frac{J_n^2 p_{\|}^2}{p_{\perp}^2}
\end{array}\right),$$
with $z \equiv k_{\perp} v_{\perp} / \Omega_{j}$, and $J_n \equiv J_n(z)$ as the $n$ th-order Bessel function. And 
\begin{equation}
U \equiv \frac{\partial f_{0 j}}{\partial p_{\perp}}+\frac{k_{\|}}{\omega}\left(v_{\perp} \frac{\partial f_{0 j}}{\partial p_{\|}}-v_{\|} \frac{\partial f_{0 j}}{\partial p_{\perp}}\right).
\end{equation}

The relationship between susceptibilities and the plasma's dielectric tensor $\boldsymbol{\epsilon}$ is as follows 
\begin{equation}
    \boldsymbol{\epsilon} = \mathbbm{1} + \sum_j \boldsymbol{\chi}_j,
\end{equation}
with the unity tensor $\mathbbm{1}$.
Finally, one can be derive the wave equation
\begin{equation} \label{alps_eq}
    \boldsymbol{n} \times (\boldsymbol{n} \times \boldsymbol{E}) + \boldsymbol{\varepsilon} \cdot \boldsymbol{E} \equiv \mathcal{D} \cdot \boldsymbol{E} = 0,
\end{equation}
where \(\boldsymbol{n} = \boldsymbol{k}c/\omega\) and $\boldsymbol{E}$ and $\boldsymbol{B}$ denote the electric and magnetic fields, respectively. 
The dispersion relation for {nontrivial} solutions ($\boldsymbol{E} \ne 0$) is finally obtained as
\begin{equation}
{\rm det}\,\mathcal{D} = 0 \label{A5}.
\end{equation}








\bibliography{bib}{}
\bibliographystyle{aasjournal}



\end{document}